\documentclass[prd,twocolumn,showpacs,preprintnumbers,amsmath,amssymb]{revtex4}

\usepackage{graphicx}
\usepackage{dcolumn}
\usepackage{bm}

\begin{document}
\preprint{HUPD-0304}

\title{Dark Energy Problem in a Four-Fermion Interaction Model}
\author{Tomohiro Inagaki\footnote{inagaki@hiroshima-u.ac.jp}}
\affiliation{%
Information Media Center, Hiroshima University, Hiroshima 739-8521, Japan
}
\author{Xinhe Meng\footnote{mengxh@public.tpt.tj.cn}}
\affiliation{%
\\
Institute of Physics, Nankai University, Tianjin, 300071, P.R.China(Post address),\\
Institute of Theoretical Physics, Academic Sinica, Beijing 100080, P.R.China
}
\author{Tsukasa Murata\footnote{murata@hiroshima-u.ac.jp}}
\affiliation{%
Department of Physics, Hiroshima University, Hiroshima 739-8521, Japan
}

\date{\today}

\begin{abstract}
Cosmology enters precision testing period with the observational
 experiments advancement. We have not arrived to decipher the origin of 
 the dominant dark energy component of our universe, although we have
 located some of its characters. 
In such a situation the possibilities of a simple four-fermion
 interaction model are investigated to describe dark energy characters. 
We explicitly calculate the vacuum expectation value of the
 energy-momentum tensor and discuss the results with comparison to the
 equation of state for dark energy requirement. It is found that the
 negative pressure is realized at low temperature. It is an appropriate
 property as a dark energy candidate. We also investigate some
 cosmological tests for the dark energy via its equation of state.
\end{abstract}
\maketitle

\section{Introduction}

Cosmological problems are as old as human history, but cosmology is a
relatively new science.
In recent years two main observational results have
revealed a profound implication in our current knowledge of the
universe.
These are the confirmation of the flat geometry and its accelerating
expansion of our universe in the relatively recent epoch, i.e., with
redshift of roughly 0.5 or less, driving by an unidentified component
as coined Dark Energy by M. Turner \cite{tn}.

The recent observations have dramatically changed our previous
intuition. The evolution of the universe was 
slowing down due to gravitational attractive force and also alleviated the
universe age crisis problem with respect to the oldest stars and
globular clusters in our Milk Way Galaxy. Respectively, the first
acoustic peak of the temperature fluctuations on the cosmic microwave
background radiation (CMBR) is observed by the
experiments of BOOMERANG, MAXIMA and WMAP \cite{mb}. It favors a spatially flat
universe and indicates our universe dominated by about 2/3 dark energy and 1/3
matter (with mostly the named non-relativistic cold dark matter) components. The
direct evidences from measurements of luminosity-redshift relations of
SN Ia \cite{SN} suggest that there exists a mysterious component so
called dark energy with a negative pressure. It is as large as  an
anti-gravity contribution to accelerate the universe 
expansion. A simple candidate which shows the properties of the dark
energy is found in the vacuum energy or cosmological constant. However,
the energy scale to the cosmological constant is about 120 orders of
magnitude smaller than that of the vacuum energy at the Planck time.
Some fine tuning of parameters is necessary in almost all the models
of particle physics. Thus a naturalness problem exists in some simple
candidates. Furthermore, recent analysis of the observational data from
WMAP combined with 2dF, SDSS, DASI and the previous CMB
observations\cite{sc} disfavors the simplest cosmological constant
explanation. 

When describing the evolution of the universe, normally we start with solving the
Einstein equation.
Since we know that recent observations favors the flat, homogeneous
and isotropic universe in Large Scale Survey, it is a good approximation
to use the energy-momentum tensor 
for the dominant matter or radiation mainly depends on time or
temperature when we consider the global evolution of the universe.
For the perfect fluid model, the energy-momentum tensor
consists of the total energy density $\rho$ and smoothly distributed
pressure $p$ by $T_{\alpha \beta}=\mbox{diag}{(\rho, p, p, p)}$.

The isotropic and homogeneous universe can be explicated by
Friedmann-Robertson-Walker (FRW) metrics. In this case Einstein
equation reduces to only two relations. One is Friedmann equation :
\begin{equation}
\frac{\dot{a}^2}{a^2}=\frac{8\pi G}{3}\rho-\frac{k}{a^2}+\frac{\Lambda}3,
\label{aa}
\end{equation}
where $a(t)$ is the scale factor in the FRW metrics.
The other is the acceleration equation
\begin{equation}
\frac{\ddot{a}}{a}=-\frac{4 \pi G }{3}(\rho+3p)+\frac{\Lambda}3.
\label{ab}
\end{equation}
These equations are very successful to describe the evolution history of
the universe from the "First Three Minutes" so far as we have observed. Furthermore, these
can accommodate the relatively recent
accelerating expansion  of our universe if we accept the existence of a
kind of vacuum energy or the cosmological constant-like term. On the other hand we
have another
solution to introduce an entity endowed with the sort of dark energy \cite{pr}.
Its energy density can govern the cosmological evolution
in Friedmann equations of standard cosmology \cite{tn}.
With the present observational data we can not finally distinguish yet whether
the accelerating expansion comes from the cosmological constant or not.

The cosmological constant can be absorbed into the energy density $\rho$
and pressure $p$ by a re-definition in the above equations.
To reconcile with 
the accelerating expansion of the universe we must have
\begin{equation}
-1\leq \frac{p}{\rho} <-\frac{1}{3} .
\label{b}
\end{equation}
The lower bound, $p/\rho = -1$, corresponds to the cosmological constant.
When we consider the dark energy model that causes the accelerating
expansion, the dominant substance has to satisfy this relation by its
own $\rho$ and $p$.

Lots of previous works are reported in concentrating on a dynamical scalar
field such as quintessence, spin-essence and so on.
In such investigations a slowly varying energy density has been
employed to mimic an effective cosmological constant \cite{rp}.
As far as we know a very little number of works have been done
in the study of the fermion field as the possible dark energy candidate.

One of the interesting candidate of the dark energy is postulated in the
model with a fermion field. In some strong coupling fermion models
dynamical symmetry breaking takes place by the fermion and the anti-fermion
condensation below a critical temperature. As is well known, the chiral
symmetry for quarks is broken down dynamically in QCD.
The composite operator of
the fermion and the anti-fermion develops the non-vanishing vacuum
expectation value. It can change behaviors of the energy density and
pressure of the universe at low temperature as the relevant energy scale
in the model can be adjusted accordingly. Therefore we launched
our plan to make a systematic study of the strong coupling fermion
models at finite temperature for a dark energy candidate.

Here we take a four-fermion interaction model as one of such
prototype models and discuss the possibilities to describe the
dark energy characters by using the thermal quantum field theory
technique \cite{bel}.
The spacetime curvature also has an important effect
on the phase structure of dynamical symmetry breaking
in the early stage of the universe
\cite{Inagaki:93ya,Rius:01dd,Inagaki:97kz}.
However, we consider the era when the main contribution
to the evaluation of the energy-momentum tensor comes from the
thermal effect. We assume that the spacetime curvature and the
expansion of the universe do not have a large influence to the
evaluation of the energy-momentum tensor.
As is known, the four-fermion interaction is non-renormalizable.
We regard the model is an effective theory which is stemming from a
more fundamental theory at a high energy scale.

In the present paper we study a four-fermion interaction model as
a dark energy candidate. By using the imaginary time formalism
we calculate the energy-momentum tensor in our model at finite
temperature. The EOS is obtained from the energy-momentum tensor.
Behaviors of the deceleration parameter are discussed to see
the dark energy property of the four-fermion interaction model.
Throughout this paper we use natural units with $c=k_B=\hbar=1$ and
assume the universe is geometric flat as astrophysical observation
favors.

\section{Energy-momentum tensor in a four-fermion interaction model}

First we briefly review the definition of the energy-momentum tensor
in the quantum field theory of fermion. The four-fermion interaction
model is one of the simplest model where dynamical symmetry breaking
takes place \cite{NJL,GN}. In $D$-dimensional spacetime it is described
by the action :
\begin{eqnarray}
S&=&\int \!d^Dx ~det(V)
\nonumber \\
&& \times
  \left[ \frac{i}2 \sum_{k=1}^{N}\left\{
      \bar \psi_k \gamma^a V^\nu_a \nabla_\nu \psi_k
  - V^\nu_a (\nabla_\nu\bar \psi_k )\gamma^a  \psi_k \right\} \right.
\nonumber \\ && \hspace{2cm} \left.
     + \frac{\lambda}{2N} \left(\sum^{N}_{k=1}
                \bar{\psi}_k\psi_k\right)^2 \right],
\label{ff}
\end{eqnarray}
where the $\lambda$ is a coupling constant, $N$ is the number of fermion
flavors and ${V_a}^\mu$ is a vierbein that relates the metric $g_{\mu\nu}$
and the local Lorentz metric $\eta_{ab}$ by
$g_{\mu\nu}=V^a_{\mu}V^b_{\nu}\eta_{ab} $.
Below we omit the flavor index $k$ of fermion field.
The action (\ref{ff})
has the global chiral symmetry under the transformation
$\psi\rightarrow e^{i\gamma_5 \theta}\psi$.
This symmetry prevent the fermion from having mass term.
Applying the auxiliary field method we rewrite the action as
\begin{eqnarray}
S&=&\int \!d^Dx ~det(V)
 \nonumber \\ && \quad \times
\left[ \frac{i}2 \left\{
    \bar \psi\gamma^a V^\nu_a \nabla_\nu \psi- V^\nu_a
             (\nabla_\nu\bar \psi )\gamma^a  \psi \right\}  \right.
 \nonumber \\ && \hspace{4cm} \left.
        - \frac{N}{2\lambda} \sigma^2 -\bar\psi\sigma\psi \right] .
\label{aC}
\end{eqnarray}
The non-vanishing vacuum expectation value of $\sigma$ denotes
the existence of the fermion and anti-fermion condensation,
$ \langle\sigma\rangle \sim -\frac{\lambda}{N}\langle\bar\psi \psi\rangle
\neq 0$. It is the signal for chiral symmetry breaking.
If the chiral symmetry broken down dynamically, the fermion
acquire a mass term proportional to $\langle\sigma\rangle$.

The classical definition of the energy-momentum tensor is \cite{BD}
\begin{equation}
  T_{\mu\nu}(x)
   = \frac{2}{[-g(x)]^{1/2}} \frac{\delta S}{\delta g^{\mu\nu}(x)}
   = \frac{V_{a(\mu}(x)}{{\rm det}[V(x)]}
        \frac{\delta S}{\delta V^{\nu)}_a(x)}.
\end{equation}
Note that the labels $a,b$ refer to the local Lorentz frame associated
with the normal coordinates, while $\mu,\nu$ denote the general
coordinate system $x^\mu$.
In the quantum field theory the expectation value of $T_{\mu\nu}$ is
defined in the sense of the path integral formalism as
\begin{eqnarray}
\Bigl<T_{\mu\nu}\Bigr>
  &=&{\frac{1}{Z}}
  \int  {\cal D}\Bigl( [-g]^{1/4}\Psi\Bigr)
        ~{\cal D}\Bigl( [-g]^{1/4}\overline{\Psi} \Bigr)
\nonumber \\
&& \times
        ~{\cal D}\Bigl( [-g]^{1/4}\sigma \Bigr)
    \frac{V_{a(\mu}(x)}{{\rm det}[V(x)] }
         \frac{\delta S}{\delta V^{\nu)}_a(x)} ~e^{i S_y},
\label{Tpath}
\end{eqnarray}
This equation is rewritten as
\begin{equation}
  \Bigl<T_{\mu\nu}\Bigr>=
  \frac{V_{a(\mu}(x)}{{\rm det}[V(x)] }{\frac{\delta}{\delta
  V^{\nu)}_a(x)}} \Bigl(-i{\rm ln} Z \Bigr).
\label{def:em}
\end{equation}
where $Z$ is the generating functional which is defined by
\begin{equation}
  Z=
  \int  {\cal D}\Bigl( [-g]^{1/4}\Psi\Bigr)
        ~{\cal D}\Bigl( [-g]^{1/4}\overline{\Psi} \Bigr)
        ~{\cal D}\Bigl( [-g]^{1/4}\sigma \Bigr)
        ~e^{i S_y}.
\end{equation}

We would like to discuss the evolution of the energy-momentum
tensor to study a fermionic dark energy model. As a first step
we confine ourselves to the Minkowski spacetime, here.
The Friedmann-Robertson-Walker metric for
the spatial flatness is conformally related to a Minkowski line
element. For the Minkowski metric the generating functional is give by
\begin{equation}
  -i{\rm ln} Z = N\Bigl\{ \Gamma[\sigma]+O(1/N)\Bigr\},
\label{def:z}
\end{equation}
where $\Gamma[\sigma]$ is the effective action :
\begin{equation}
  \Gamma[\sigma]
  =\int\! d^D x \left\{ \frac{-\sigma^2}{2\lambda_0}
  -i {\rm Tr}~{\rm ln} (i\gamma^\mu \partial_\mu-\sigma) \right\}.
\label{def:eac}
\end{equation}
The energy-momentum tensor $(\ref{def:em})$ contains a zero-point
vacuum energy and is divergent. We use the point splitting
prescription to regularize the expectation value of it \cite{BD}.
Substituting the Eqs.(\ref{def:z}) and (\ref{def:eac})
into Eq.(\ref{def:em}) we obtain
\begin{equation}
  \Bigl<T_{\mu\nu}(x)\Bigr>
  = \eta_{\mu\nu}\frac{N}{2\lambda_0}\sigma^2
  -i N \lim_{x^\prime\rightarrow x}{\rm Tr}~\gamma_{(\mu}\partial_{\nu)}^x
  S_F(x,x^\prime;\sigma),
\label{tmunu}
\end{equation}
in the leading order of $1/N$-expansion
where $\partial^x$ denote the differentiation with respect to $x$ and
$S_{F}(x,x^\prime;\sigma)$ is the Feynman propagator for the free fermion
with mass $\sigma$ which satisfies
\begin{equation}
  (i\gamma^\alpha\partial_\alpha-\sigma)S_{F}(x,x^\prime;\sigma)
     =  \delta^D(x,x^\prime).
\end{equation}
The second term in the right hand side of Eq.(\ref{tmunu}) corresponds to
the expectation value of the energy-momentum tensor of the free fermion with
mass $\sigma$. Therefore the energy-momentum tensor of the four-fermion
interaction model is given by the sum of the quadratic potential term of
$\sigma$, i.e., the first
term in the right hand side of Eq.(\ref{tmunu}), and the energy-momentum
tensor of the massive free fermion in the leading order of $1/N$-expansion.
It should be noted that the fermion
mass $\sigma$ is a dynamical variable. We can determine the dynamically
generated fermion mass by evaluating the effective potential.

\section{Energy-momentum tensor at finite temperature}

Next we study the thermal effect to the energy-momentum tensor.
As is shown in Eq.(\ref{tmunu}), the vacuum expectation value
for the energy-momentum tensor is described by the free fermion
propagator.
Following the imaginary time formalism, the fermion propagator
at finite temperature is obtained from the one at $T=0$ by the
Wick rotation and the replacements \cite{bel, Wein}
\begin{eqnarray}
   \int dk^0 &\rightarrow & \frac{1}{\beta}\sum_n ,\\
   k^0 &\rightarrow & i \omega_n,\\
   \gamma^0 &\rightarrow & -i\gamma^4 ,
\end{eqnarray}
where $\beta=1/T$. The discrete variable $\omega_n$ is
given by
\begin{equation}
\omega_n\equiv i\frac{2n+1}{\beta}\pi,
\end{equation}
according to
the anti-periodic boundary condition of the fermion field.
After these replacement and some calculations the energy-momentum
tensor (\ref{tmunu}) at finite temperature in the leading order
of the $1/N$ expansion reads
\begin{eqnarray}
&& \langle T^{00}\rangle = N\sigma^2/{2\lambda}\nonumber \\
&&\quad
-\frac{2Ntr1}{(4\pi)^{(D-1)/2}\Gamma\left(\frac{D-1}{2}\right)}
 \int^\infty_0 dK K^{(D-2)}
\nonumber\\ && \hspace{1.2cm} \times
   \left[ \frac{\sqrt{K^2+\sigma^2}}2
     \tanh{\left(\frac{\beta}2\sqrt{K^2+\sigma^2} \right)}
       - \sum \frac{1}{\beta} \right]\nonumber\\
&& \hspace{1.2cm} -\langle 0|T^{00}|0\rangle ,
\label{C}
\end{eqnarray}
and
\begin{eqnarray}
&&\langle T^{ii}\rangle  = -N\sigma^2/{2\lambda}  \nonumber \\
&& \qquad -\frac{2Ntr1}{(D-1)(4\pi)^{(D-1)/2}
      \Gamma\left(\frac{D-1}{2}\right)}
          \int^\infty_0 dK K^{(D-2)} \nonumber \\
&& \hspace{1.2cm} \times
 \left[\frac{K^2}{ 2\sqrt{K^2+\sigma^2}}
    \tanh{\left(\frac{\beta}2\sqrt{K^2+\sigma^2} \right) } \right]\nonumber \\
&& \hspace{1.2cm} - \langle 0|T^{ii}|0 \rangle ,
\label{C2}
\end{eqnarray}
where $tr1$ is the trace with respect to the fermion legs
in an arbitrary dimensions $D$, $tr1=2^{D/2}$.
Since the zero-point of energy-momentum tensor is unknown parameter
in the quantum field theory, we subtract the expectation value
for $T=0$, $\langle 0|T^{\mu \nu}|0 \rangle$, to normalize it.
Since the four fermion interaction is not renormalizable,
we consider the spacetime dimension less than $4$ to
regularize the theory. Here we regards the theory for
$D=4-2\epsilon$ with $\epsilon$ sufficiently small positive
as a regularization of the one in four dimensions.
There is a correspondence between the positive regularization
parameter $\epsilon$ and the cut-off scale $\Lambda$ of the
theory \cite{Inagaki:97kz},
\begin{equation}
\frac{1}{\epsilon}-\gamma+\ln 4\pi +1
\leftrightarrow
\ln \left(\frac{\Lambda}{\mu}\right)^2,
\end{equation}
where $\mu$ is a renormalization scale.

\begin{figure}
\includegraphics[width=6.8cm]{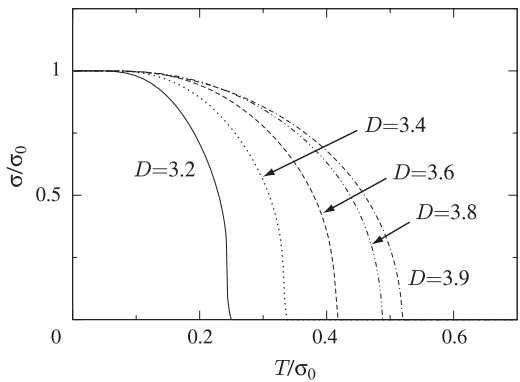}
\caption{Behavior of the expectation value of $\sigma$ for 
$\lambda = -20 \sigma_0^{2-D}$ as a function of $T$. 
$\sigma_0$ is the expectation value at $T=0$.}
\label{fig:sigma}
\end{figure}

To obtain the expectation value of $\sigma$ which corresponds to
the dynamical fermion mass we evaluate the effective potential
$V(\sigma)$ of the four-fermion interaction model at finite temperature.
In arbitrary spacetime dimensions $2 \leq D < 4$ it is given by \cite{IKM}
\begin{eqnarray}
V(\sigma)&=& \sigma^2/{2\lambda}
\nonumber \\
&& -\frac{1}{\beta}
\frac{2 tr1}{(4\pi)^{(D-1)/2}\Gamma\left(\frac{D-1}{2}\right)}
\nonumber \\
&& \qquad \times \int dK K^{D-2} \left[ \ln \cosh
   \left( \frac{\beta}{2}\sqrt{K^2+\sigma^2} \right) \right.
\nonumber \\
&& \hspace{2.2cm} \left. -\ln \cosh\left( \frac{\beta}{2}K \right) \right] ,
\end{eqnarray}
at the large $N$ limit. The expectation value of $\sigma$ is
determined by observing the minimum of the effective potential. In the
present paper we are interested in the influence of the chiral symmetry
breaking. Hence we take the coupling $\lambda$ larger than the critical
one and numerically calculate the behavior of $\sigma$ as is shown
in Fig.~\ref{fig:sigma}.
The dynamical fermion mass is generated below a critical
temperature, $T_c$. Details of the generated fermion mass is disscussed
in Ref. \cite{IKM} for $2\leq D<4$.

Inserting the expectation value of $\sigma$ into the Eqs.
(\ref{C}) and (\ref{C2}) we obtain the energy-momentum tensor
of the four-fermion interaction model at finite $T$.

For the case of chiral symmetry restoration (i.e. $T < T_c$)
the fermion is massless. We get the traceless energy-momentum tensor
\begin{equation}
T_{\alpha \beta}=\mbox{diag}(\rho, \rho/(D-1), \rho/(D-1), \rho/(D-1), \cdots),
\label{es}
\end{equation}
where the density $\rho$ is introduced by
\begin{eqnarray}
&&\rho =
 \frac{Ntr1}{(4\pi)^{(D-1)/2} \Gamma\left(\frac{D-1}{2}\right)}\nonumber\\
&&
 \times\int_0^\infty dK \frac{K^D}{\sqrt{K^2+\sigma^2}}
     \left\{1-\tanh{\left(\frac{\beta}2\sqrt{K^2+\sigma^2}
                \right)} \right\} ,
\nonumber \\
\end{eqnarray}
which is always positive by numerical calculations and
formally satisfies the photon like equation of state at high temperature,
$w=p/\rho=1/(D-1)$, with $D=4-2\epsilon$.
The behavior for $w=p/\rho=1/(D-1)$ is very reasonable because $\sigma =0$
means the massless field model! So its equation of state must be similar
to the photon's.

For the case of chiral symmetry breaking with $\sigma$ nonzero, we have
\begin{eqnarray}
T_{\alpha \beta}&=&\rho_0 \eta_{\alpha \beta}
\nonumber \\ && \quad
 + \mbox{diag}(\rho, \rho/(D-1), \rho/(D-1), \cdots)
\nonumber \\ && \qquad
 +\mbox{diag}(co(\sigma),0,0, \cdots),
\label{es1}
\end{eqnarray}
where the small correction
\begin{eqnarray}
co(\sigma)&=&
 \frac{Ntr1}{(4\pi)^{(D-1)/2} \Gamma\left(\frac{D-1}{2}\right)}\nonumber\\
&& \quad
 \times\int_0^\infty dK \frac{K^{D-2} \sigma^2}{\sqrt{K^2+\sigma^2}}
\nonumber \\
&& \quad
  \times \left\{1-\tanh{\left(\frac\beta2 \sqrt{K^2+\sigma^2} \right)}\right\} ,
\end{eqnarray}
is also positive and the energy-momentum tensor elements also
satisfy the EOS for the dark energy with the background of temperature
dependent cosmological constant
$\rho_0=N[\sigma(T=0)^2-\sigma(T)^2]/(2\lambda)$.

\begin{figure}
\includegraphics[width=6.8cm]{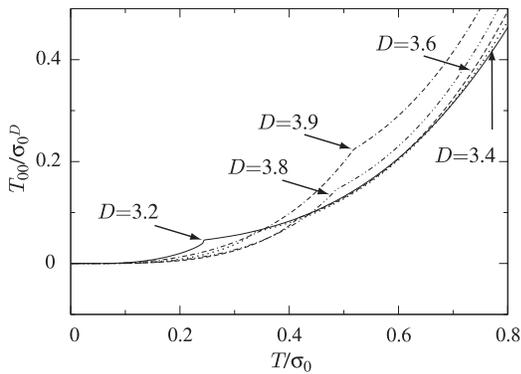}
\caption{Behavior of the energy-momentum tensor $T_{00}$ 
for $\lambda = -20 \sigma_0^{2-D}$ as a function of $T$.} 
\label{fig:T00}
\end{figure}

\begin{figure}
\includegraphics[width=6.8cm]{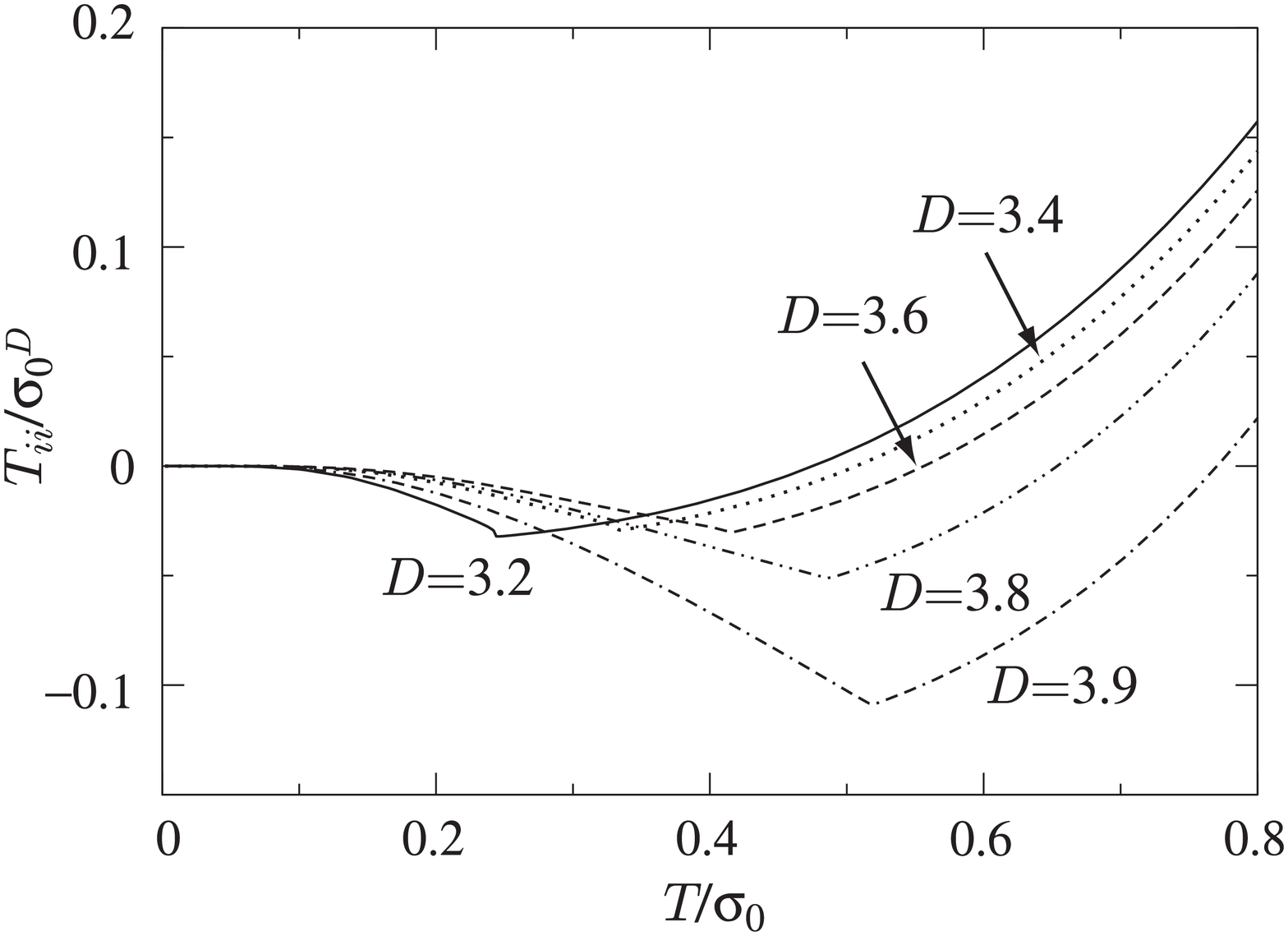}
\caption{Behavior of the energy-momentum tensor $T_{ii}$ 
for $\lambda = -20 \sigma_0^{2-D}$ as a function of $T$.} 
\label{fig:Tii}
\end{figure}

The behavior of the energy-momentum tensor is illustrated in Figs.
\ref{fig:T00} and \ref{fig:Tii}. As is explained above, zero-point of
$\langle T^{\mu\nu} \rangle$ can not be determined in the quantum field theory.
We normalize it which satisfies
$\langle T^{\mu\nu} \rangle = 0$ at $T=0$.
As is clearly seen, the negative pressure is realized for a low
temperature region. The lines turn sharply at the critical temperature $T_c$.

In the following section we turn to the the classical
cosmological tests of EOS for our model.

\section{Dynamical evolution in the FRW cosmology}

In addition to the Friedmann equation we have the important equation to
satisfy when we solve the evolution of the universe. It is the conservation
for the energy-momentum tensor of the cosmic fluid given by
\begin{equation}
\dot{\rho}_i+3H(p_i+\rho_i)=0 ,
\label{ag}
\end{equation}
where subscript $i$ describes each species of the fluids in the
considering universe. Here we mainly discuss the evolution of the
universe which is dominated by the dark energy of our model.

Since the recent observational data strongly suggest that our universe
is spatially flat \cite{pryke, netterfield, stompor}, that is $k=0$
in Eq.(\ref{aa}), we will also investigate only this case here.
In this framework we mainly discuss several classical observational
tests \cite{ka} for this mysterious dark energy problem.

A dimensionless quantity is more useful to discuss physical phenomena
of our model. In the standard cosmology, for
convenience, the density fraction of the world is expressed
in terms of a critical density which is defined by the total energy
density, i.e. $\rho_c=3H_0^2/8\pi G$.
When we treat the flat, homogeneous and isotropic universe,
the comoving coordinates is convenient to describe the dynamical
evolution of it. In this treatment the coordinates at any place do
not vary with the evolution of the universe. It is described by a
scale factor $a(t)$ that is a function of only cosmic time in the
homogeneous and isotropic universe. If we
find the relationship between the scale factor and the cosmic time,
then the evolution of the universe is clearly understood. For
example, we can directly deduce the Hubble parameter and the
deceleration parameters as a function of time, the dependence
of the luminosity distance on redshift and so on.
By investigating its dynamical evolution in the dark energy dominant
era, we compare the observable quantities derived in our model with
the experimental data.

We first consider a general case where equation of state is written by
\begin{equation}
p=w\rho ,
\label{ea1}
\end{equation}
The parameter $w$ depend on the spacetime dimension $D$ and this dependence
is assumed to be extremely small for low temperature region.
By using the continuity equation (\ref{ag}) and the Eq.(\ref{ea1}),
we easily obtain the evolution of the energy density $\rho$ which obeys
\begin{equation}
\rho=\rho_0a^{-n},
\label{eb}
\end{equation}
where $\rho_0$ is the present energy density of the universe and $n$
is given by $n=3(1+w)$. Substituting
Eq.(\ref{eb}) in the Friedmann equation, we get
\begin{equation}
 H^2 = H_0^2\left[
        \Omega_t a^{-n} + \Omega_M a^{-D+1}
           +\Omega_R a^{-D}\right] ,
\label{ec}
\end{equation}
where $H_0$ is the Hubble parameter today and $\Omega_a$ is the density
parameter defined by $\Omega_a=\rho_{a0}/\rho_c$.
Since we consider the flat universe, the
density parameters must satisfy the relation, $\sum_i \Omega_i =1$.

It is hard to solve the Friedmann equation (\ref{ec}) in general
situations, because we do not know the EOS for the dark energy. From
Eq.(\ref{ec}), however, we can easily obtain the evolution of
the scale factor. If we consider the dark energy dominant situation,
the scale factor behaves as
\begin{eqnarray}
a(t) &=& \left (
\frac{\sqrt{n^2H^2_0\Omega_t}}{2}t\right)^{2/n} ,
 \qquad {\rm for} \quad n>0
\nonumber \\
a(t) &\propto& \exp \left(\sqrt{ H^2_0\Omega_t }t \right) .
 \qquad {\rm for} \quad  n=0
\label{ee}
\end{eqnarray}
The latter solution shows the same behavior with the
inflationally expansion when the energy scale is sufficiently high.
We may use it to describe the accelerating expansion with a
cosmological constant in the low energy scale, as observed today.

\begin{figure}
\includegraphics[width=6.8cm]{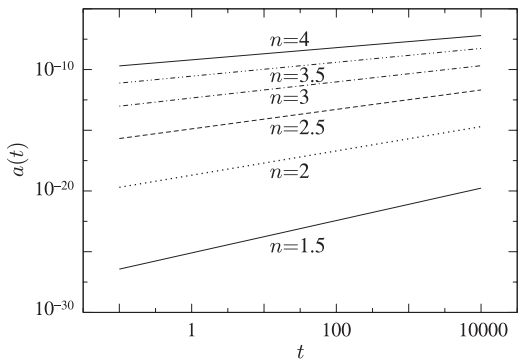}
\caption{This figure shows the scale factor of Eq.(\ref{ee}) as a
 function of cosmic time. Here we use the following values $\Omega_t=0.75 $,
 $H_0=71{\rm km/s\cdot Mpc}$. }
\label{scale}
\end{figure}

The dynamical evolution factor is just
the function of EOS which is shown in Fig.~\ref{scale}. Comparing the
lines in the figure, we can see that for the smaller value of $n$
the scale factor $a(t)$ increases faster.

\section {The deceleration parameter}

Following the fact that our universe is expanding, one may ask a
question that the expansion of our universe is accelerating or
decelerating now, or how fast it expands?
A parameter to describe the evolution of the universe is not only
the Hubble parameter but also another important parameter, so
called the deceleration parameter $q$ which is defined by
\begin{equation}
q=-\frac{a\ddot{a}}{\dot{a}^2}.
\label{fa}
\end{equation}
The parameter $q$ is dimensionless parameter which describes the
rate for slowing the expansion. By using Eq.(\ref{ee}) we obtain
\begin{equation}
q = \frac12 (n-2) = \frac12 (3w +1) .
\label{fb}
\end{equation}
As a first consideration we assume that the state equation
parameter $w$ is constant in the dark energy dominant era.

\begin{figure}
\includegraphics[width=6.8cm]{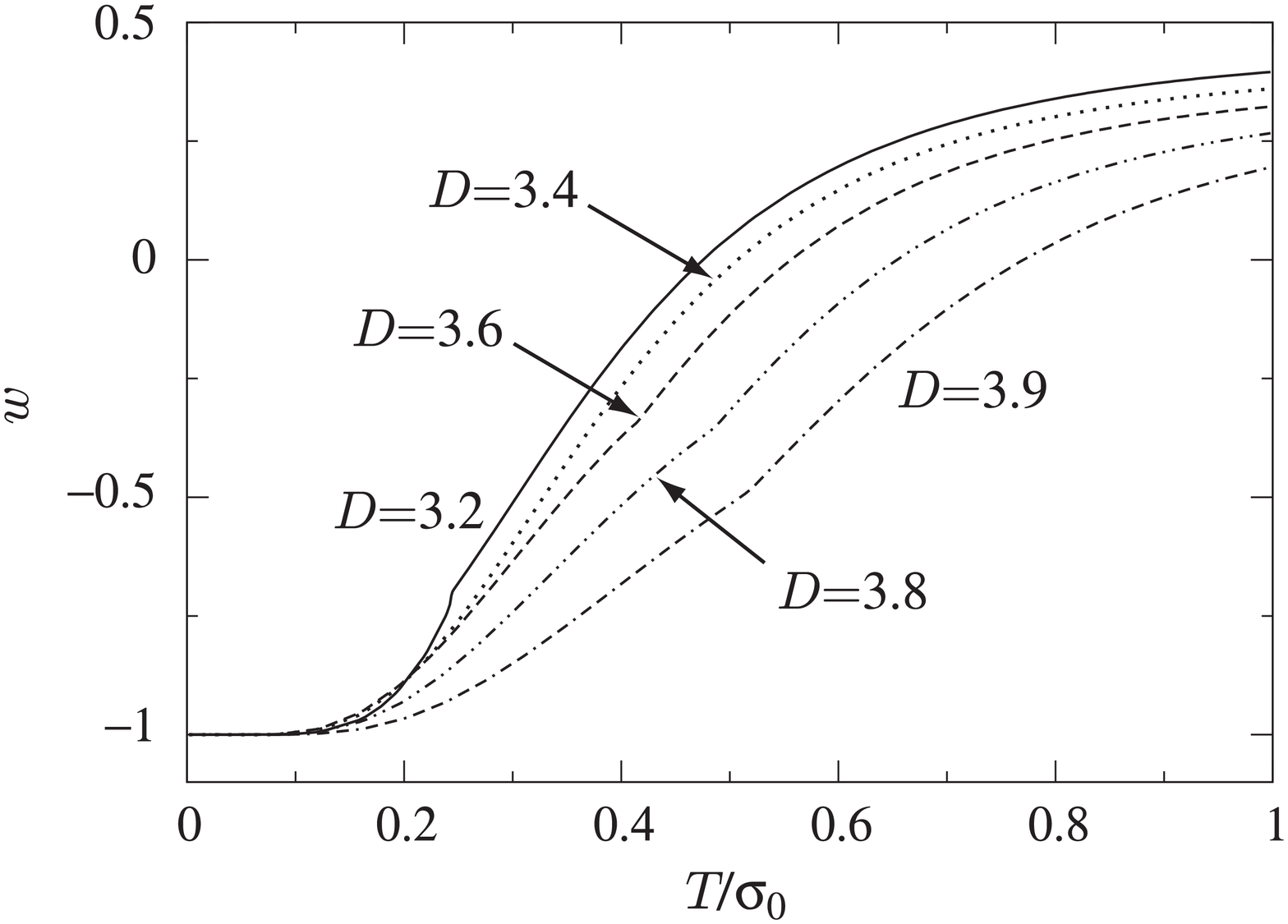}
\caption{The evolution of the state equation parameter $w$
with respect to $T/\sigma_0$. We set $\lambda = -20 \sigma_0^{2-D}$. }
\label{fig:w}
\end{figure}

In the four-fermion interaction model we show the evolution of the
deceleration parameter in Fig.~\ref{fig:w}.
As is clearly seen in Fig.~\ref{fig:w} $w$ is not a constant. The
evolutions of $w$ depend on both the temperature and the regularization
parameter $\epsilon$ which is related to the cut-off scale of the theory.
However, we think that the constant $w$ is a good approximation when we
consider the present evolution of our universe because our universe
is bathed in enough low temperature now. 
Generally speaking,
the deceleration parameter $q$ is a function of both the present density
$\Omega_i$ and $w$. Furthermore, it is also a function of the
parameter $\epsilon$ because we consider a non-renormalizable model
as a low energy effective theory.
Figure \ref{fig:w} shows that the relationship between the parameter
$w$ and the regularization parameter $\epsilon$. The present $w$
(low temperature region) seems almost a monotone function of the
parameter. The parameter $w$ smoothly falls down and the curves become
flat as $T$ decreases. At the lower temperature limit $w$ approaches
the cosmological constant limit. It shows the exponential expansion
of the present universe.  As is shown in Eq. (\ref{es}) $w$ becomes
$1/(D-1)$ at the high temperature limit. It is a property of the
massless fermion. Therefore the exponential expansion only appears
at the low temperature region in our model without any fine tunings.
It is a very important property to a dark energy candidate. Slopes of
each lines in Fig.{\ref{fig:w}} are depend on the coupling
$\lambda$. For larger coupling $\lambda$ the slope becomes
steeper. However, the behaviors of $w$ does not change qualitatively as
$\lambda$ varies. 

From the recent SN Ia and Cosmic Microwave Background
observations \cite{pr} a cosmological constant-like dark energy may exist
today which accelerate the expansion of the universe. It suggests that
the decelerating parameter lies in a range, $-1< q < 0$.
This observational constraint is naturally satisfied in the four-fermion
interaction model. Recently there are many dark
energy models of simple scalar fields like the quintessence \cite{rp},
however, they all need fine tuning of some parameters to certain extents. In the present
model, on the other hand, fine tuning is not necessary. In other words
the model has some properties which are realized in the quintessence
scenario without any cosmological constant-like terms. Certainly we need
more SN Ia observational data to confirm the important constraint, which
is just the main mission of the near future SNAP project.

\section{Concluding Remarks}

We investigated the cosmological effects of the vacuum
expectation value of energy momentum tensor in a four fermion
interaction model. According to the imaginary time formalism we derive
the explicit form of the energy momentum tensor.
The energy-momentum tensor depends on the temperature and the
dynamically generated fermion mass. The fermion field is massless
and has the similar property as the radiation field above the
critical temperature. We consider that the temperature of the present
universe is much less than the critical one and the chiral symmetry
is broken down dynamically. Thus the composite operator of the fermion
and the anti-fermion develops a non-vanishing vacuum expectation
value in the present stage of the universe. The vacuum expectation value
plays a role of the cosmological constant-like.

Applying it to models of the universe without an explicit cosmological
constant term, we discussed the dependence of the scale factor $a(t)$
on the state equation parameter $w$ and evaluate the deceleration
parameter $q$.
We found that the behavior of the EOS is dramatically changed at low
temperature. A dark energy property was naturally obtained in the
four-fermion interaction model.
In the present analysis the fermion mass scale can be at the scale
of neutrino, QCD, electroweak symmetry breaking, extra-dimension
\cite{abe} and so on. We have had a loosely constrained parameter
range for the cosmological constant \cite{sw}.
For more concrete analysis we must apply the results to some
cosmological tests with comparisons to observational data,
which are under our investigations.

Our model is only a prototype one, but we believe that the results
are worth discussing further about the real world. By observational
data analysis we know the dark energy is almost smoothly distributed
\cite{av}, which implies a kind of
very weak interaction that we are not very clear about yet. Here we just
employ the simple four-fermion
interaction model to give the global energy momentum tensor which
depends on temperature and some model parameters. It might be too
simple, however we found interesting results. As a next step,
it is very important to apply our analysis to other well-defined
and more realistic models. One of the interesting models is SUSY NJL model
\cite{Buchmuller:1982ty,Buchbinder:1997ta}.
For the SUSY NJL model the effective action is written as
\begin{eqnarray}
\Gamma[ \sigma]
    &=& \int d^4x \left[ {\rm Tr}\ln(i\not{\!\partial}-\sigma) \right.
\nonumber \\ &&\qquad \left.
       -2 {\rm Tr}\ln(\partial^2- \sigma^2) +N\sigma^2/(2\lambda) \right].
\label{db}
\end{eqnarray}
According to the supersymmetry, contributions from the fermion
and boson loops cancel out each other exactly at $T=0$. Since
the thermal effect restore the chiral symmetry, dynamical symmetry
breaking does not take place even at finite temperature.
If we introduce a soft SUSY breaking term such as
$\bigtriangleup \phi^* \phi$ and/or $\bigtriangleup \phi^{c*} \phi^c$
in the action \cite{Hashida}, then the contribution from the boson loop
is modified. The composite operator $\bar{\psi} \psi$ can develop
a non-vanishing vacuum expectation value which satisfies the dark energy
property to some range.
The scalar mass depends on the coupling parameter. When
the coupling parameter is large enough, $\bigtriangleup > 2 \sigma$,
the scalar mass $m=\sqrt{(\sigma^2-\bigtriangleup)}$ becomes heavier
than the dynamical fermion mass.
We can choose the coupling parameter $\bigtriangleup$ to
realize the appropriate scale for the cosmological constant
and the negative pressure of the present universe accordingly.

So far the origin of the dark energy in the current cosmological
models is still under discussions with many theoretical speculations.
It is accommodated in the standard Friedmann cosmology
based on the relativistic gravitational theory with a cosmological
constant-like term. It can also be interpreted in extended
gravity scenarios and corresponding cosmological models, like the brane cosmologys.
That attracts us to further research works continuously.

\begin{acknowledgments}

The paper is started in X.~H.~Meng's academic visit at Hiroshima
University supported from JSPS INVITATION Fellowship.
The authors have also benefited a lot from discussions with many people,
especially, D.~Kimura, G.~Senjanovich, X.~P.~Wu, K.~Yamamoto and Z.~H.~Zhu
. This work is also supported partly by grant of No.NK-BRSF
G19990753 from Ministry of Science and Technology of China.

\end{acknowledgments}

\end{document}